\documentclass[%
	 reprint,
	preprintnumbers,
	nofootinbib,
	 amsmath,amssymb,
	 aps,
	floatfix,
	multicol,
	]{revtex4-1}
	
\usepackage[italicdiff]{physics}

\usepackage{graphicx,color,hyperref}
\usepackage{dcolumn}
\usepackage{bm}
\usepackage{subfigure}

\usepackage{ulem}

\newcommand{\Dz}{\Delta z}

\usepackage{listings}
\begin{document}

\preprint{OU-HET 1058}

\title{Deep Learning and AdS/QCD}

\author{Tetsuya Akutagawa}
\email{akutagawa@het.phys.osaka-u.ac.jp}
\author{Koji Hashimoto}%
\email{koji@phys.osaka-u.ac.jp}
\author{Takayuki Sumimoto}
\email{t\_sumimoto@het.phys.osaka-u.ac.jp}
\affiliation{%
 Department of Physics, Osaka University, \\
  Toyonaka, Osaka 560-0043, Japan
}%

\begin{abstract}

We propose a deep learning method to build an AdS/QCD model from the data of hadron spectra.
A major problem of generic AdS/QCD models is that a large ambiguity is allowed for
the bulk gravity metric with which QCD observables are holographically calculated.
We adopt the experimentally measured spectra of $\rho$ and $a_2$ mesons
as training data, and perform a supervised machine learning which determines concretely
a bulk metric and a dilaton profile of an AdS/QCD model.
Our deep learning (DL) architecture is based on the AdS/DL correspondence \cite{Hashimoto:2018ftp}
where the deep neural network is identified with the emergent bulk spacetime.

\vspace*{10mm}

\end{abstract}

\maketitle


\section{Introduction}
\label{sec:intro}

The AdS/CFT correspondence
\cite{Maldacena:1997re,Gubser:1998bc,Witten:1998qj}, 
or the holographic principle, is a promising way to define a quantum gravity.
In spite of its importance, a fatal problem is to find a dual gravity system for a given QFT, for which so far no
systematic approach has been successful. Particularly important QFTs are those with Yang-Mills sectors, including QCD,
whose large $N$ and strong coupling limit are believed to give a classical gravity dual, while we are lacking
in how to construct  the dual concretely and explicitly. So far, we know necessary conditions for the gravity dual,
such as symmetries and spectral properties, as well as recently investigated OTOCs 
\cite{Shenker:2013pqa,Shenker:2013yza,Maldacena:2015waa} and computational complexities \cite{Susskind:2014rva}.

The string theory ``top-down'' construction does not solve the problem, because it merely provides examples 
as a pair of a gravity and a QFT at the same time. 
Many excellent work provided a pair in which the QFT side resembles a given target QFT.
A lot of effort has been put to seek for a gravity dual of QCD, as QCD is the renowned, and realistic QFT among all.

It should be emphasized that once a classical gravity system is given, the dual QFT quantities can be easily calculated
by the AdS/CFT dictionary. The problem is how we can go backwards: for a given QFT correlators, how we get the gravity system.
To solve this kind of inverse problems, we need special techniques. In particular, strongly coupled QFT consists of
a vast amount of data, such as $n$-point correlators for infinite kinds of local/nonlocal gauge invariant operators.
Furthermore, QCD is a part of the Standard Model for which a lot of experimental data is available.
To this end, machine learning method may help us. If there exists a gravity dual of QCD which is simple enough
with a finite number of parameters, the features of the data of QCD needs to be efficiently extracted, by solving the AdS/CFT
backwards.  Deep learning \cite{Hinton,Bengio,LeCun}
which technically advanced these years may shed light on the fatal problem of the AdS/CFT.

In Ref.~\cite{Hashimoto:2018ftp}, 
deep learning was applied to determine an emergent gravity metric from a given data of a QFT.
The dictionary used was for a one-point function of an operator in the QFT, corresponding to a bulk field value in
the dual gravity system. In Ref.~\cite{Hashimoto:2018bnb}, the method was applied to lattice QCD data of a chiral condensate, to find 
a metric of the gravity system dual to QCD. Based on the success of this method working for one-point functions,
in this paper we make one more step toward realistic QCD. We use hadronic two-point functions, {\it i.e.}
the hadron spectra which are measured in experiments.

Needless to say, the most well-observed quantities in experiments for QCD are hadron spectra and hadron couplings.
The best framework to test the deep learning method is the AdS/QCD \cite{Erlich:2005qh,DaRold:2005mxj,Karch:2006pv}, a bottom-up construction of phenomenological
gravity models based on symmetries and the dictionary. It is known that the simple AdS/QCD framework can host lots of
QCD quantities including hadron spectra.
Here, again, the conventional methods in the AdS/QCD is
first to come up with a gravity model, then to calculate the QCD quantities from the model and then to compare those
with experimental data. If the quantity does 
not match well, one throws away the gravity model and try again with a different gravity model.
The gravity model has a large arbitrariness, and in addition the dictionary is nonlocal, so solving the inverse problem is
challenging. This is the reason why the deep learning method can help finding the gravity dual of QCD.

The AdS/DL correspondence in Ref.~\cite{Hashimoto:2018ftp} 
was invented due to the similarity between the deep neural network (DNN) and bulk gravity system.\footnote{
For a detailed relation, see Ref.~\cite{Hashimoto:2019bih}. 
An early study on the similarity between the AdS/CFT and the DL is in  Ref.~\cite{You:2017guh}. 
See Refs.~\cite{Gan:2017nyt,Lee:2017skk} for related essays. A continuum limit of the deep layers was studied in a different 
context \cite{deepest}.}
In the training, weights of the neural network are trained and determined by machine, which is regarded as an emergence of
the spacetime, as the differential equation on the discretized gravity spacetime is regarded as a propagation of information
on the deep neural network, with the depth direction identified with the AdS radial direction.

In this paper, we upgrade the deep neural network given in Refs.~\cite{Hashimoto:2018ftp,Hashimoto:2018bnb} 
to accommodate hadron two-point functions
as a supervised learning, and use the experimental data of the mass of $\rho$ mesons and $a_2$ mesons at zero temperature 
as the 
training data. The neural network is a discretized bulk action of 
the AdS/QCD model \cite{Erlich:2005qh,DaRold:2005mxj,Karch:2006pv}, with the metric
and the dilaton fields identified as the network weights.
With some physically reasonable regularizations, the supervised learning of the neural network is successful, ending
up with smooth profiles of the gravity metric and the dilaton configuration in 5 spacetime dimensions, as the trained weight parameters of the neural network.
Namely, the inverse problem of finding the gravity system for a given hadron spectra is solved by the deep learning.
Using the obtained metric and dilaton, we can predict excited hadron masses which were not used as the training data.

The advantage of the deep learning method is that it can provide a systematic approach to determine the
gravity dual from a given dataset of the QFT correlators, which even generalizes for prediction.
Using the data analysis methods of deep learning \cite{Carleo:2019ptp}, 
we may hope to combine all possible information of QCD
as the training data to discover a proper gravity dual.
For that, various QCD requirements studied in improved holographic QCD \cite{Gursoy:2007cb,Gursoy:2007er}
will help.

This paper is organized as follows.
First, in Sec.~\ref{sec:AdS/QCD} we make a brief review of the concept of the 
renowned soft-wall AdS/QCD model of Ref.~\cite{Karch:2006pv}. Based on the model,
in Sec.~\ref{sec:NN} we construct our deep neural network by discretizing the
bulk equation of motion.\footnote{While finalizing this article, 
we noticed a paper dealing with the shear viscosity 
\cite{Yan:2020wcd} in which a similar discretized neural network is used.} 
We provide our deep learning architecture with experimental
vector meson mass spectra as our supervising dataset, and describe our hyperparameters
and regularizations.
The deep learning is performed in Sec.~\ref{sec:Opt}. After a reproduction test of the background
of Ref.~\cite{Karch:2006pv}, we use the experimentally measured data to train our AdS/QCD model
to find an optimized, emergent background geometry and dilaton. 
We discuss physical implication of our emergent geometry.
Sec.~\ref{sec:summary} is for our conclusion.
App.~\ref{Sec:app} includes the details of our deep learning architecture.


\section{Review: the AdS/QCD model}
\label{sec:AdS/QCD}

The AdS/QCD  \cite{Erlich:2005qh,DaRold:2005mxj,Karch:2006pv}
is to provide a simple phenomenological model in 5-dimensional curved spacetime which describes
desired sectors of QCD effectively.
The 5-dimensional Lagrangian is built under the guide of the AdS/CFT dictionary. For example, 
to compute hadron mass spectra, one introduces fields propagating in a 5-dimensional curved spacetime 
which are dual to the hadrons of concern. 
One of the most popular models of AdS/QCD is so-called soft-wall models given in Ref.~\cite{Karch:2006pv}
where, rather than using a brute cut-off of the 5-dimensional geometry as in the inaugural work 
\cite{Erlich:2005qh,DaRold:2005mxj},
one introduces a 5-dimensional dilaton field to realize a smooth wall to confine fields in the curved geometry,
which enables discussions on a part of the QCD Regge trajectories.

In this section, we briefly review the model given by Ref.~\cite{Karch:2006pv} on which our deep learning architecture is built.  
With the explicitly given 5-dimensional action, the vector meson spectra can be calculated by the model.

Generically, any AdS/QCD model is given in the following manner. 
First of all, one assumes the existence of an effective theory which is holographically dual to QCD via the AdS/CFT correspondence. 
Next, based on the well-known dictionary of the AdS/CFT, one writes the 5-dimensional bulk action of the theory, with ingredients 
necessary to reproduce  a QCD sector of one's concern. For example, for flavor symmetry, one introduces corresponding gauge symmetry in the bulk theory. The associated gauge bosons in 5 dimensions correspond to the flavor current operators in QCD, which are nothing but the vector mesons. 
The gauge bosons are in a 5-dimensional curved spacetime whose gravity metric and bulk dilaton field are prearranged.
	
Let us focus on the vector meson spectra in QCD.
The gravity side is the 5-dimensional effective action of a $U(1)$ gauge theory
	\begin{align} \label{eq:effective action of vector sector}
		I_\text{vector} &= - \frac{1}{4g_5^2} \int dz d^4x e^{-\Phi(z)} \sqrt{-g} F_{MN} F^{MN} \, , 
	\end{align}
where $F_{MN}$ is the field strength of the 5-dimensional massless gauge field $V_M(z,x^\mu)$, 
and $g_5$ is the gauge coupling constant. The indices $M,N$ represent those of the 5-dimensional spacetime, 
while the directions along the 4 dimensions associated with QCD are denoted as $\mu,\nu$. 
The emergent 5-th direction is parameterized by the coordinate $z$ $(\geq 0)$. Everything is made dimensionless
by using the AdS radius $L$.

The theory is in a curved spacetime. The soft wall models include the gravity metric $g_{MN}(z)$ and the dilaton field 
$\Phi(z)$ as the background fields. In particular, the dilaton field is essential in the spectral analyses in Ref.~\cite{Karch:2006pv}. 
The metric is written with a single function $A(z)$ without losing its generality,
	\begin{align} \label{eq:metric}
		g_{MN}dx^Mdx^N
		&= e^{2A(z)} \qty(dz^2 + \eta_{\mu\nu} dx^\mu dx^\nu) \, , 
	\end{align}
because QCD is in an infinitely extended flat Lorentzian spacetime, at zero temperature.
For the use of the AdS/CFT correspondence, it is assumed that the curved spacetime is asymptotically AdS: 
$A(z) \sim -\log z$ near the AdS boundary $z =0$, where the unit $L=1$ is used.
	
Following Ref.~\cite{Karch:2006pv}, we choose a gauge $V^z = 0, \partial_\mu V^\mu=0$. 
Using the plane wave basis for the 4 dimensions, 
we consider a solution of the form 
	\begin{align}
		V^\mu(z,x^\mu)=v(z)c^\mu e^{-ikx} \, ,
	\end{align} 
with a mass-shell condition $-k^2 = m^2$ for the 4-dimensional mass $m$. 
Then we obtain the following equation which the coefficient function $v(z)$ should satisfy,
	\begin{align} \label{eq:eom of bulk field}
		\dv{z} \qty( e^{-B(z)} \dv{z} v(z) ) +m^2 e^{-B(z)} v(z) = 0 \, .
	\end{align}
Here we have defined the combination
	\begin{align} \label{eq:background of soft wall}
		B(z) \equiv \Phi(z) - A(z) \, .
	\end{align}

In the AdS/CFT dictionary, for the current operator of QCD to be excited, the corresponding modes in the gravity side
need to be normalizable. 
The differential equation \eqref{eq:eom of bulk field} has normalizable solutions only for discrete values of the mass 
$m = m_n$ ($n=0,1,2,\cdots$). 
Therefore, these discrete values $m_n$ are interpreted as the vector meson spectrum.

\begin{figure}[t]
	\centering
	\includegraphics[width=240pt]{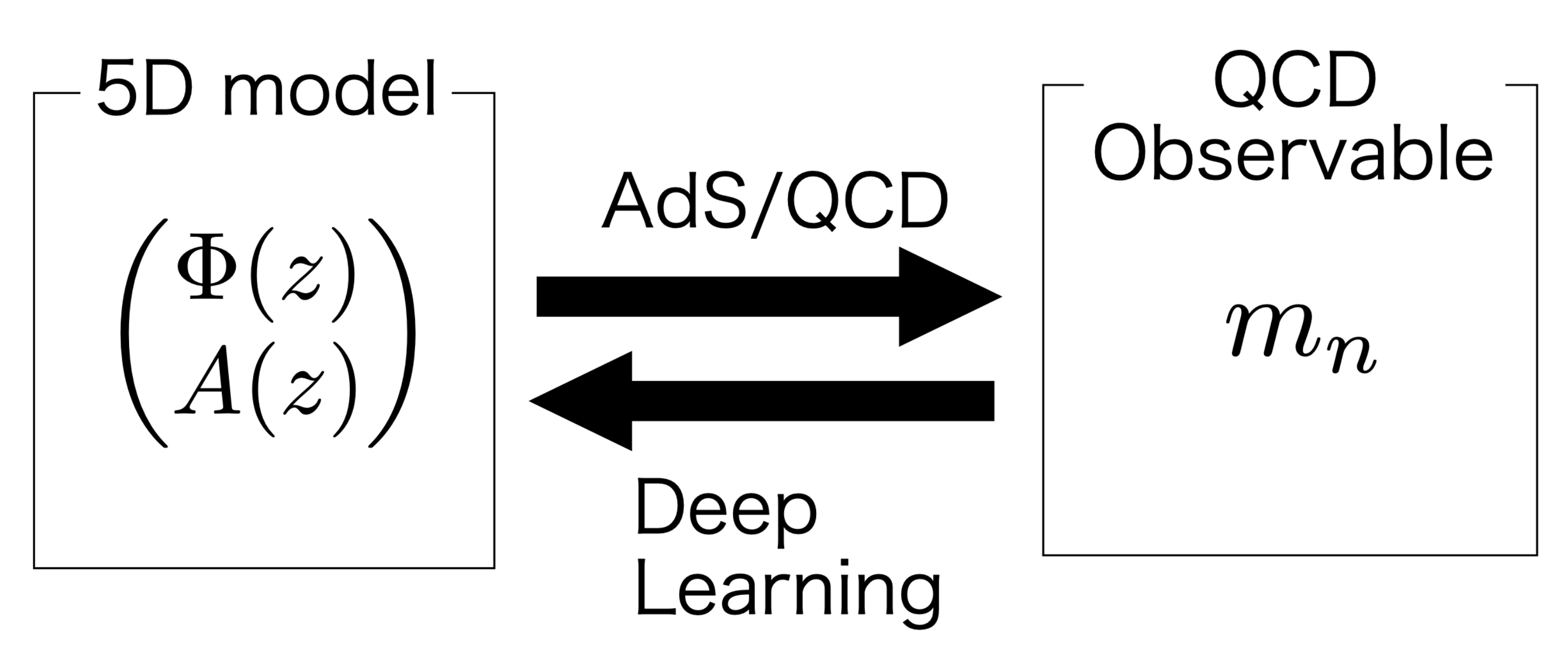}
	\caption{In the ordinary AdS/QCD modeling, one prepares the background 
	metric and dilaton fields ($A(z)$ and $\Phi(z)$) and 
	then calculates the QCD observables (the hadron mass $m_n$). 
	The judgement of whether the chosen background is appropriate is checked after matching the calculated observables with experimental data. 
	On the other hand, our AdS/DL approach solves backward.}
	\label{fig:illastration_of_AdSQCD}
\end{figure}

So, summarizing the procedures, once the explicit form of the metric function $A(z)$ and the dilaton profile $\Phi(z)$
is given, one can calculate the normalizable solutions of \eqref{eq:eom of bulk field} to obtain the vector meson mass spectra.
See Fig.~\ref{fig:illastration_of_AdSQCD} for the illustration. In Ref.~\cite{Karch:2006pv} the following functions are used,
	\begin{align}\label{KarchAnsatz}
  		A(z)=-\log z \, , \quad \Phi(z) = z^2 \, .
	\end{align}
This ansatz is simple enough since $A(z)=-\log z$ means that the whole 
5-dimensional spacetime is $AdS_5$, and $\Phi(z)=z^2$
was adopted to produce the asymptotic Regge behavior in the hadron spectra.
	
One can also calculate the spectra of higher spin mesons from the 5-dimensional model. 
A spin-$S$ meson is dual to a rank-$S$ symmetric tensor field \cite{Katz:2005ir}. 
By the procedures similar to those of the vector meson case, 
the $z$-dependent part of the bulk tensor field obeys the same equation as \eqref{eq:eom of bulk field} with a
different definition of $B(z)$;
	\begin{align} \label{eq:spin dependence of B}
		& B(z) = \Phi(z) -\qty(2S-1)A(z).
	\end{align}
One understands that the spin dependence of the meson spectra is encoded in the background bulk field profiles.

Now, one can see that 
the difficulty of the AdS/QCD model building is in solving inversely the duality:
the direction of the standard procedures is only from the gravity to QCD quantities. One needs the explicit metric and the dilaton
which are not known a priori.\footnote{One can of course resort to Einstein dilaton equations of motion assumed, to constrain
possible profiles of the metric and the dilaton fields, see Refs.~\cite{Gursoy:2007cb,Gursoy:2007er}. 
A partial list of the obstacles are
(i) dilaton potential allows arbitrariness, (ii) since QCD is $N=3$ quantum gravity corrections such as higher derivatives in curvatures may exist, and (iii) string theory effective action is too restrictive at tree level while generic quantum gravity corrections are unknown.  } 

On the other hand, this paper aims at ``optimizing'' the model by experimental data. More concretely, we determine the model background $B(z)$ by using deep learning of the training data of hadron masses. 
See Fig.~\ref{fig:illastration_of_AdSQCD}.
In addition, using the spin dependence of the definition of $B(z)$ in \eqref{eq:spin dependence of B}, we can obtain the spacetime metric and the dilaton separately: the machine finds different $B(z)$'s from the data for the mesons with $S=1$ and $S=2$ respectively, 
then the metric and the dilaton are obtained as 
	\begin{align} \label{eq:metric and dilaton}
		& A(z) = \qty(B_{S=1} - B_{S=2})/2  \, , \\
		& \Phi(z) = \qty(3B_{S=1} - B_{S=2})/2 \, .
	\end{align}
Using these, one can calculate other QCD observables as a prediction of the determined model.
Our proposal offers a new data-driven approach to the AdS/QCD model building.

	\begin{figure*}[t]
		\centering
		\includegraphics[width=14cm]{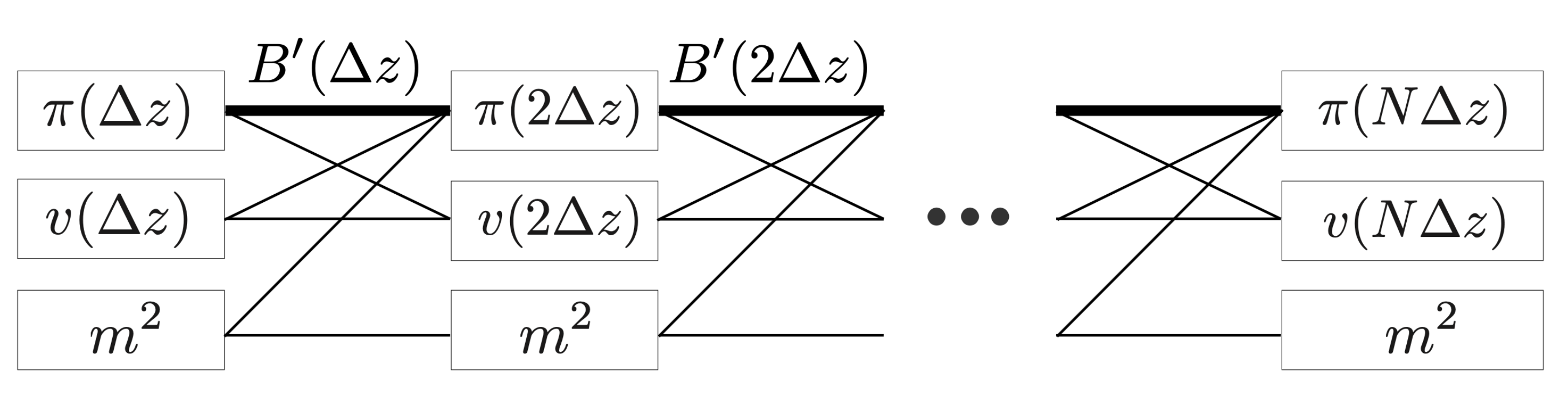}
		\caption{
			Neural network representation of \eqref{eq:defference eq}. 
			The layer depth direction (horizontal direction in this figure) corresponds to the emergent 
			radial direction of the 5-dimensional spacetime. The input is on the left, the output is on the right.
			Only the thick lines are trainable weights, 
			corresponding to
			the bulk background field $B'(z)$. The binary classification layer at the output is not
			shown in this figure, for simplicity.
		}
		\label{fig:illustration_DNN}
	\end{figure*}


\section{Neural Network for AdS/QCD}
\label{sec:NN}

In this section we construct our neural network. According to the concept of the 
AdS/DL correspondence \cite{Hashimoto:2018ftp}, 
we present a deep neural network architecture optimizing a generic soft-wall AdS/QCD model.
The optimization is conducted under a supervised deep learning, with respect to the QCD experimental data,
where the meson spectrum $m$ in \eqref{eq:eom of bulk field} is treated as the input data. 

The implementation can be divided into three steps. 
First, we translate the equation of motion (EOM) of the bulk field into a deep neural network. The network itself is regarded as a spacetime, and the metric on the discretized spacetime corresponds to trainable weight parameters in the neural network. 
The next step is to arrange the input and the output layers for a binary classification. The input data is the vector meson spectrum data, and the output data is related to the normalizability condition of the solution of the bulk field equation.  
The remaining step is technical for the training to be successful:  fixing hyperparameters and introducing a regularization. 
We have to adjust the value of the hyperparameters such as the discretization spacing and the number of layers. 
We also need a regularization which come from physical requirements such as smoothness of the spacetime.
In the following, in each subsection we explain the three steps described above.

\subsection{Deep neural network}
\label{Sec:DNN}

We prepare a deep neural network where trainable weight parameters are identified with the configuration of 
the gravity/dilaton fields of the holographic model. 
We identify the radial propagation \eqref{eq:eom of bulk field} of the bulk vector field from $z$ to $z+\Dz$ 
with the propagation of the information on the neural network from one layer to another.

We introduce a conjugate momentum field 
	\begin{align} \label{eq:canonical variable}
		\pi(z) \equiv \pdv{z} v(z)
	\end{align}
to reduce \eqref{eq:eom of bulk field} to a set of first order differential equations,
one of which is
	\begin{align} \label{eq:canonical form of eom}
		0
		= & \, \pdv{z} \qty\Big( e^{-B(z)} \pi(z) ) + m^2 e^{-B(z)} v(z)  \nonumber \\
		= & \, e^{-B(z)}\left[
		-B'(z) \pi(z) + \pi'(z) + m^2 v(z)\right] \, .
	\end{align}
Then we discretize the $z$ coordinate in \eqref{eq:canonical variable} and \eqref{eq:canonical form of eom} with a spacing 
$\Dz$,
	\begin{align} \label{eq:defference eq}
		\begin{cases}
			\pi(z+\Delta z) &= \pi(z) + \Delta z \, (B'(z) \pi(z) - m^2 v(z)) \\
			v(z+\Delta z) &= v(z) + \Delta z \, \pi(z)
		\end{cases}
	\end{align}
The derivatives acting on the vector field $\pi$ and $v$ are replaced by the difference, and we leave the form
$B'(z)\equiv\pdv*{B}{z}$ as it is. 

With this \eqref{eq:defference eq}, 
let us make a neural network representation of the bulk field equation.  
The propagation equation \eqref{eq:defference eq} tells us that 
the values of $(\pi,v)$ at $z=N\Dz$ with an integer $N$ can be computed from the initial values of $(\pi,v)$ at
$z=\Dz$ and $m$, when a background $B'(z)$ is given. Namely, the output of the system is given by the input and $B'(z)$,
	\begin{align} \label{eq:deference eq formal expression}
		\mqty(\pi(N\Delta z) \\	v(N\Delta z))
		= F\qty\Big(\pi(\Delta z),	v(\Delta z),m;B'(z)) \, .
	\end{align}
Recall that in a generic feedforward deep neural network (DNN) 
the unit values at the final layer are calculated by the unit values at the 
initial layer with a given trainable weights. The concept of the AdS/DL correspondence is based on this similarity 
\cite{Hashimoto:2018ftp}. Following that, we find the dictionary between the bulk field equation and 
the neural network as shown in TABLE~\ref{tab:dictionary of EOM/DNN}.

	\begin{table}[bht]
		\centering
		\begin{tabular}{c|c}
			EOM & DNN  \\ \hline \hline
			$z$ & layer label \\
			$B'(z)$ & weights \\
			$\pi(z), v(z), m$ & units \\
			$F$ & architecture
		\end{tabular}
		\caption{The relationship of between the EOM \eqref{eq:canonical form of eom} and 
		the DNN. Here $m$ is $z$-independent 
		in \eqref{eq:canonical form of eom}, so it corresponds to a unit of a fixed value.}
		\label{tab:dictionary of EOM/DNN}
	\end{table}

Equation \eqref{eq:deference eq formal expression} is nothing but the deep neural network itself, with the
layer depth direction identified as the emergent radial direction $z$. See FIG.~\ref{fig:illustration_DNN}.
The AdS boundary $z=0$ is
identified with the input layer, while the deep infrared region of the emergent space $z=\infty$ is
the output layer.
Generically, neural network weights are matrix-valued trainable parameters, while our weights include only $B'(z)$ 
as the trainable parameters. 
This means that our neural network is sparse, and most of weight components are put to 0 or some fixed value.

As we mentioned, the 5-dimensional spacetime is asymptotically AdS${}_5$ near $z=0$, 
so we can determine the values of $(\pi, v)$ at the initial layer $z=\Dz$, based on their behavior in the AdS.
Using the asymptotic solution of \eqref{eq:eom of bulk field} with the AdS metric (and also 
assuming that the dilaton vanishes there), we find 
	\begin{align}
		\pi(\Dz) = 2S(\Dz)^{2S-1} \, , \quad v(\Dz) = (\Dz)^{2S}\, .  \label{eq:BC on initial layer}
	\end{align}
See App.~\ref{app:initiallayer} for the derivation. 
The overall proportionality magnitude of $v$ and $\pi$ is ignored since the EOM \eqref{eq:eom of bulk field} is linear in $v(z)$.
See FIG.~\ref{fig:illustration_DNN} for the whole schematic view of
our neural network.


\subsection{Dataset for binary classification}

The next step is to prepare training dataset for our supervised leaning. Since we want to extract possible features from the experimental data of the meson spectrum, we may simply use the value of the spectrum as our input. 
Then, what should the output data be? 
In view of  \eqref{eq:deference eq formal expression}, the output data is the values of the vector field at $z=\infty$.
Depending on whether the mode $v(z)$ is normalizable or not, the values of $v(z)$ at large $z$ vanishes or diverges.
It needs to be normalizable only when we have the correct value of the input mass $m$.
In this way, we can arrange the neural network as a binary classification problem, by putting a discrimination label
to the input mass.
	
More concretely, as the input data we prepare a set of random real numbers (the mass).
The range of the generated random numbers include the experimentally measured values of the meson mass. 
If the random number is close to (far from) the experimental values in the real spectrum, it is named as a positive (negative) data.
In the training data, the output value 
as the label for the positive (negative) data should be 0 (1), which is the discrimination label.
Therefore, we give our neural network a task to classify the input real numbers: a meson mass classifier.

Since we expect the final output to be 0 or 1, we introduce an additional layer there.
This layer outputs 0 or 1 according to whether the value of $v(N\Delta z)$ satisfies the following normalizability 
condition or not,
	\begin{align} \label{eq:discrimination}
		v(N\Delta z) e^{-B(N \Dz)/2} \le \epsilon \, .
	\end{align}
For this, we arrange as the final layer a smeared box function of the width $\epsilon$, with no weight multiplication.\footnote{
Here, $\pi$ also needs to be normalizable for a positive data, but in our implementation we do not look at it.
This is because $\pi$ is expected not to converge well at large $z$, due to discretization errors.}
And as for the loss function, we simply adopt L1 loss.

	
\subsection{Hyperparameters and regularization}
\label{Sec:hyp}

For the machine learning to work, we have to tune the hyperparameters in the architecture, and also need to
introduce regularization terms to the loss function.

Our hyperparameters are, the number of layers $N$, and the discretization spacing $\Dz$ which appears in
the fixed weights of the neural network,\footnote{There is an option to treat $\Delta z$ as additional trainable parameters. 
	This $\Delta z$ could even depend on $z$. 
	This training of the $z$-dependent spacing $\Delta z$ appears to be a quantum gravity, which is intriguing, 
	and we leave the optimization of the spacing in AdS/DL to a future problem.}
and the discrimination threshold $\epsilon$ in \eqref{eq:discrimination} in the classification layer.
In fact, these are closely related to each other from a physical viewpoint, as follows.

First, $N$ and $\Dz$ give the infrared location $z=N\Dz$
at which whether the vector field $v(z)$ behaves as a normalizable function or not is verified, while
the threshold for the discrimination is $\epsilon$. Although in principle we have $\epsilon \to 0$ for $N\to\infty$, 
for our numerical calculations we need a finite $N$. In addition, even though the analytic solution goes to 0 at $z\to\infty$,
due to the discretization error of the differential equation \eqref{eq:eom of bulk field} the value at $z=N\Dz$ could deviate 
from 0. So, for the network to work properly even with the discretization effect, we need to tune $\epsilon$ such
that it also allows the deviation. 
In practice, we fix these hyperparameters by using the analytic solutions in the famous soft wall model of 
Ref.~\cite{Karch:2006pv}, see App.~\ref{app:hyperparameters} for the details.

The regularization which should be added to the loss function is introduced by the following three reasons.
The fist one is the spacetime interpretability. The obtained set of weights is interpreted as $B(z)$, which is the dilaton field 
and the metric field. They need to be a smooth function of $z$, otherwise there is no physical interpretation 
\cite{Hashimoto:2018ftp,Hashimoto:2018bnb}. Second, as in \eqref{eq:BC on initial layer}, for the AdS/CFT to work,
we impose the asymptotic AdS condition, which constrains the form of $B(z)$ near the initial layer, the 
boundary $z=0$.
The third reason is the soft wall. For \eqref{eq:eom of bulk field} of the vector field to 
have normalizable solutions,\footnote{The EOM
\eqref{eq:eom of bulk field} can be rewritten in the form of a Schr\"odinger equation (see Ref.~\cite{Karch:2006pv}), 
and the potential of the Schr\"odinger equation depends only on derivatives of $B(z)$. In the model of Ref.~\cite{Karch:2006pv} 
the background \eqref{KarchAnsatz} corresponds to a harmonic oscillator potential at large $z$, which is the ``wall'' of
a confining geometry. 
We introduce a regularization so that the machine can find a similar behavior.}
the background field needs to have an infrared wall. This is also related to 
the technical trainability of the neural network, and to the choice of the random
initial configurations of $B'(z)$ before the training.

Therefore, we introduce the following three kinds of regularization terms:
	\begin{itemize}
		\item $B(z)$ is a smooth function of $z$.
		\item $B(z)$ is asymptotically AdS at small $z$.
		\item $B(z)$ has a ``wall'' at large $z$.
	\end{itemize}
For concrete functional forms of these regularizations in the loss function, see App.~\ref{app:loss}.


\section{Optimization by the Data of Meson Spectra}
\label{sec:Opt}

In this section, we determine the background of the AdS/QCD model \eqref{eq:effective action of vector sector} by training our neural network with the dataset of the meson spectrum. We have two numerical experiments.
The first one is a test case to check whether our deep learning can actually reproduce the soft wall model \eqref{KarchAnsatz}
of Ref.~\cite{Karch:2006pv} from the data which is generated by the model with \eqref{KarchAnsatz} in advance. 
We confirm that our architecture can learn the model successfully.
The second numerical experiment is
the optimization of the model by deep learning with the experimental data of the meson spectrum. 
The machine finds a metric function $A(z)$ and a dilaton profile $\Phi(z)$ which are consistent with the experimental data.
We discuss physical implication of the determined AdS/QCD model.

		\begin{figure}[t]
			\centering
			\subfigure[]{
				\includegraphics[width=0.8\linewidth]{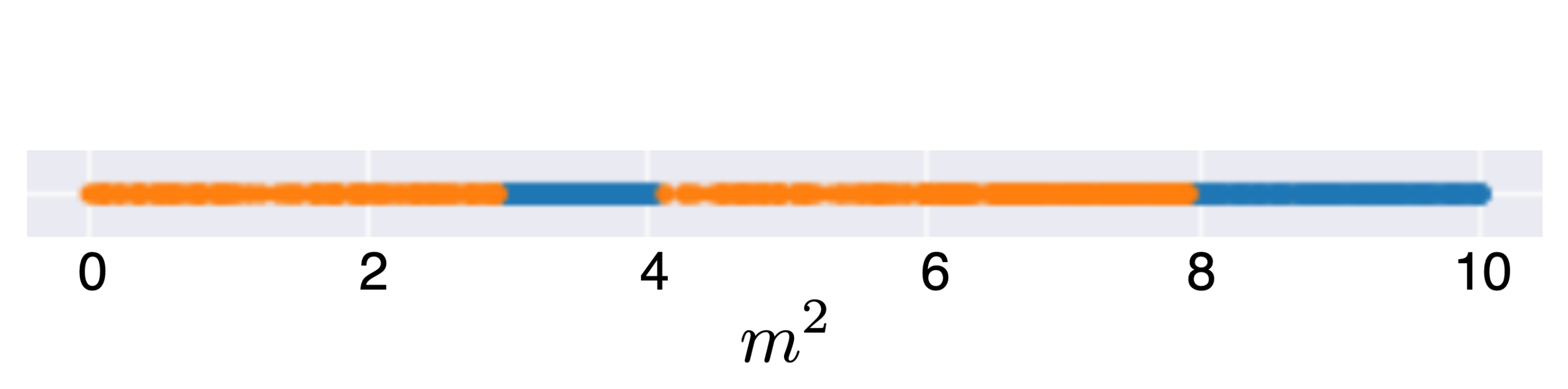} \label{fig:dataset_test}
			}
			\subfigure[]{
				\includegraphics[width=0.8\linewidth]{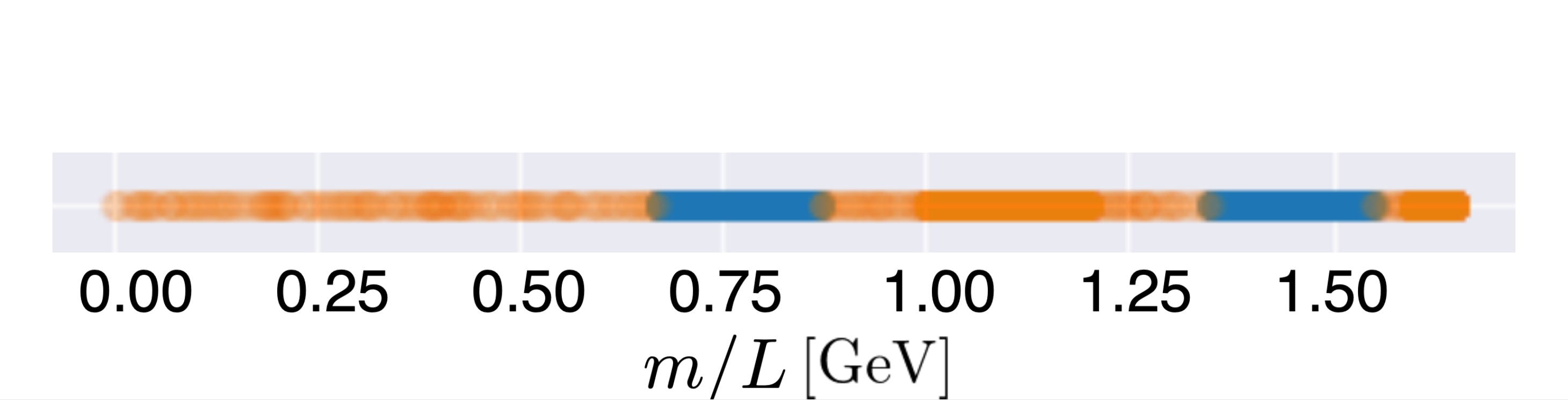} \label{fig:dataset_rho}
			}
			\subfigure[]{
				\includegraphics[width=0.8\linewidth]{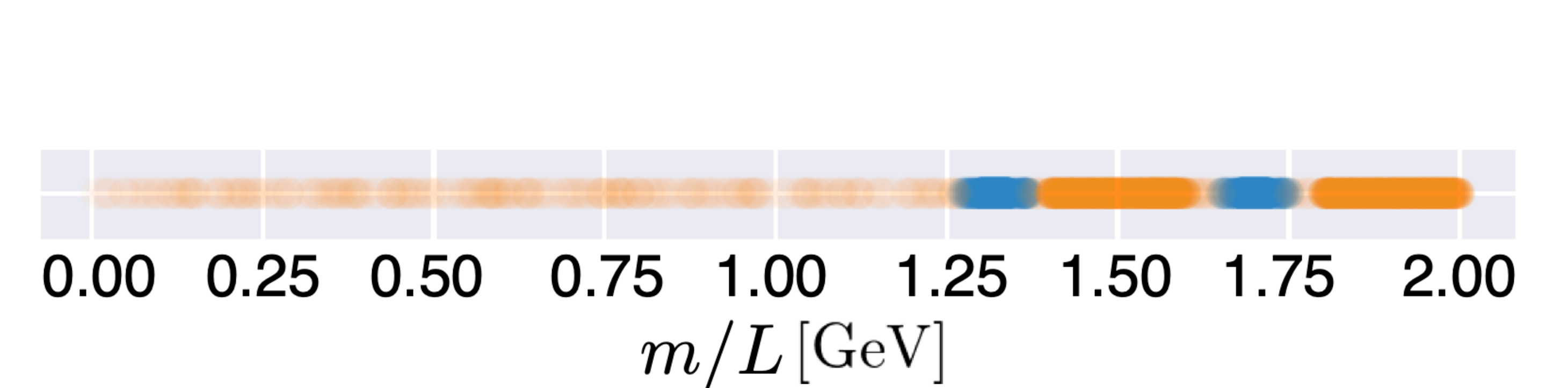} \label{fig:dataset_a2}
			}
			\caption{
			Visualization of the positive/negative datasets we use. Blue/orange points represent positive/negative data. 
			The data (a) is for the spectrum $m^2$ of the model of Ref.~\cite{Karch:2006pv} used in Sec.~\ref{sec:test}, 
			and the data (b) and (c) are
			for the spectra $m/L$ of the $\rho$ meson and for $a_2$ meson used in Sec.~\ref{sec:exp}, respectively.
			}
			\label{fig:dataset}
		\end{figure}


\subsection{Reproduction test}
\label{sec:test}

Since the implementation of our neural network is totally based on the generic soft wall model,
as a first check of whether our architecture works properly, we perform a reproduction test of the model of 
Ref.~\cite{Karch:2006pv} with \eqref{KarchAnsatz}.
First, we prepare the mass spectra calculated by the model with \eqref{KarchAnsatz}. Then, using only that set of data,
we train our neural network to find $B'(z)$. Then we compare the function $B'(z)$ which machine determined, and
the function \eqref{KarchAnsatz}. If we confirm that they are similar enough, then we claim that our architecture works
and the model of Ref.~\cite{Karch:2006pv} is reproduced.

For the dataset of $m^2$, 
we generate a set of real numbers in the range $[0,10)$. This range includes the first two levels of the spectrum,
since the meson masses calculated by the model of Ref.~\cite{Karch:2006pv} is $m_n^2=4(n+1)$  with $n=0,1,2,\cdots$
(see Fig.~\ref{fig:dataset_test}). 
Note that the widths of the region of the positive data (the blue dots in Fig.~\ref{fig:dataset_test}) are chosen by hand,
which are hyperparameters.

\begin{figure}[t]
			\centering
			\includegraphics[width=240pt]{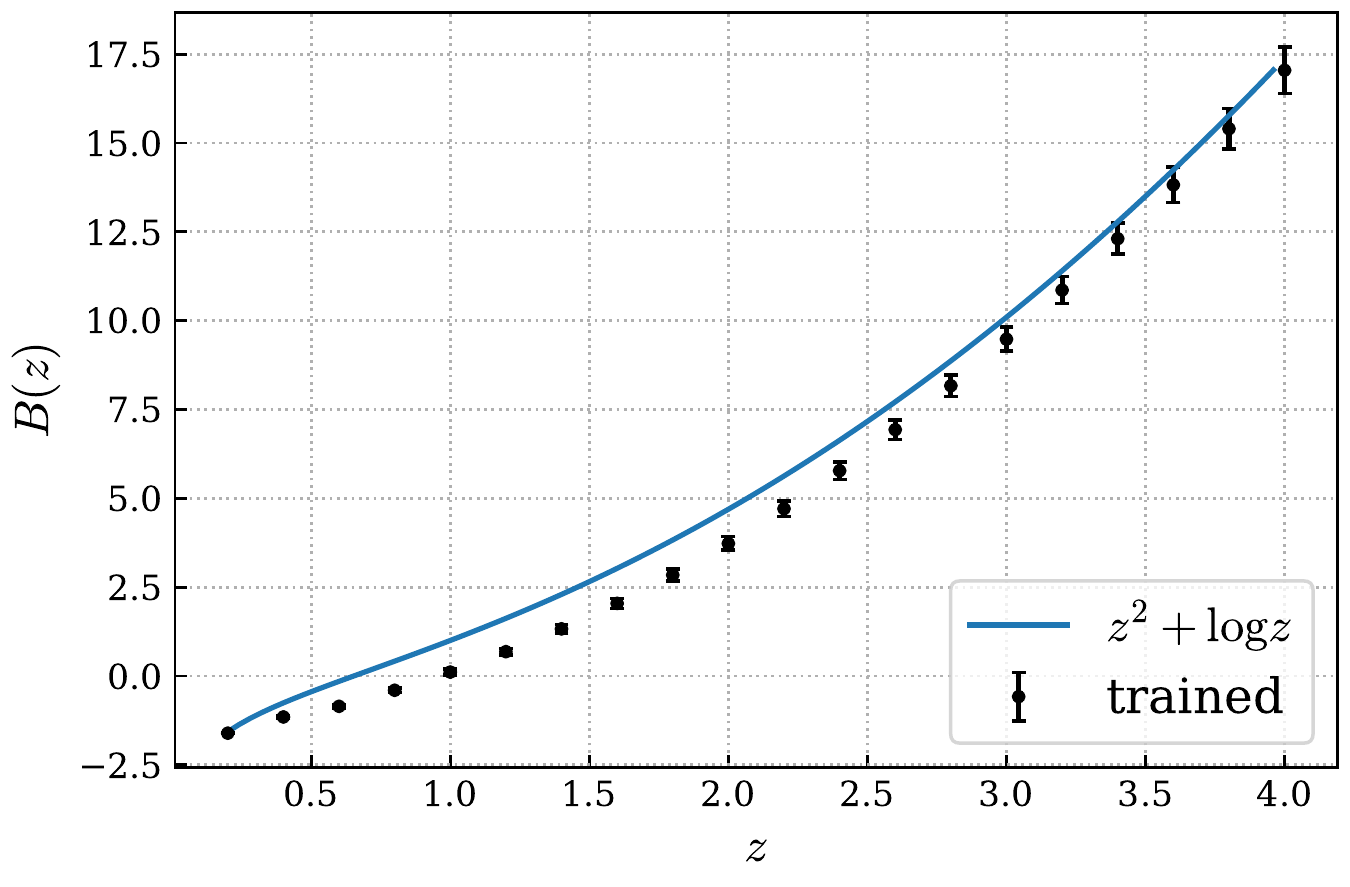}
			\caption{The trained result of the reproduction test, $B(z)$. The trained 20 functions $B'(z)$ (for 20 trainings) 
			are averaged, then discretely integrated over $z$ to get $B(z)$, plotted with 2$\sigma$ error bars. 
			Blue curve represents the model of Ref.~\cite{Karch:2006pv}, $B(z) = \log z + z^2$ 
			given in \eqref{KarchAnsatz}.}
			\label{fig:soft wall model}
		\end{figure}

In FIG.~\ref{fig:soft wall model}, we show the result of our deep learning with $N=20$, $\Dz=0.2$, $\epsilon=0.25$,
and 20 times of the repetition of the training.
The result shows that the soft wall model of Ref.~\cite{Karch:2006pv} is reproduced well, qualitatively. 
Although we introduce some physical regularizations such as the asymptotic AdS condition and the wall condition,
it is worth noted that only a part of the meson spectra is used for training the model.
Hence we claim that our deep learning architecture works 
for the meson spectrum as the input data. 
And in particular, the AdS/DL paradigm is shown to be helpful to construct effective AdS/QCD models.

		\begin{figure*}[tbh]
			  \centering
			\subfigure[]
			 {\includegraphics[scale=0.6]{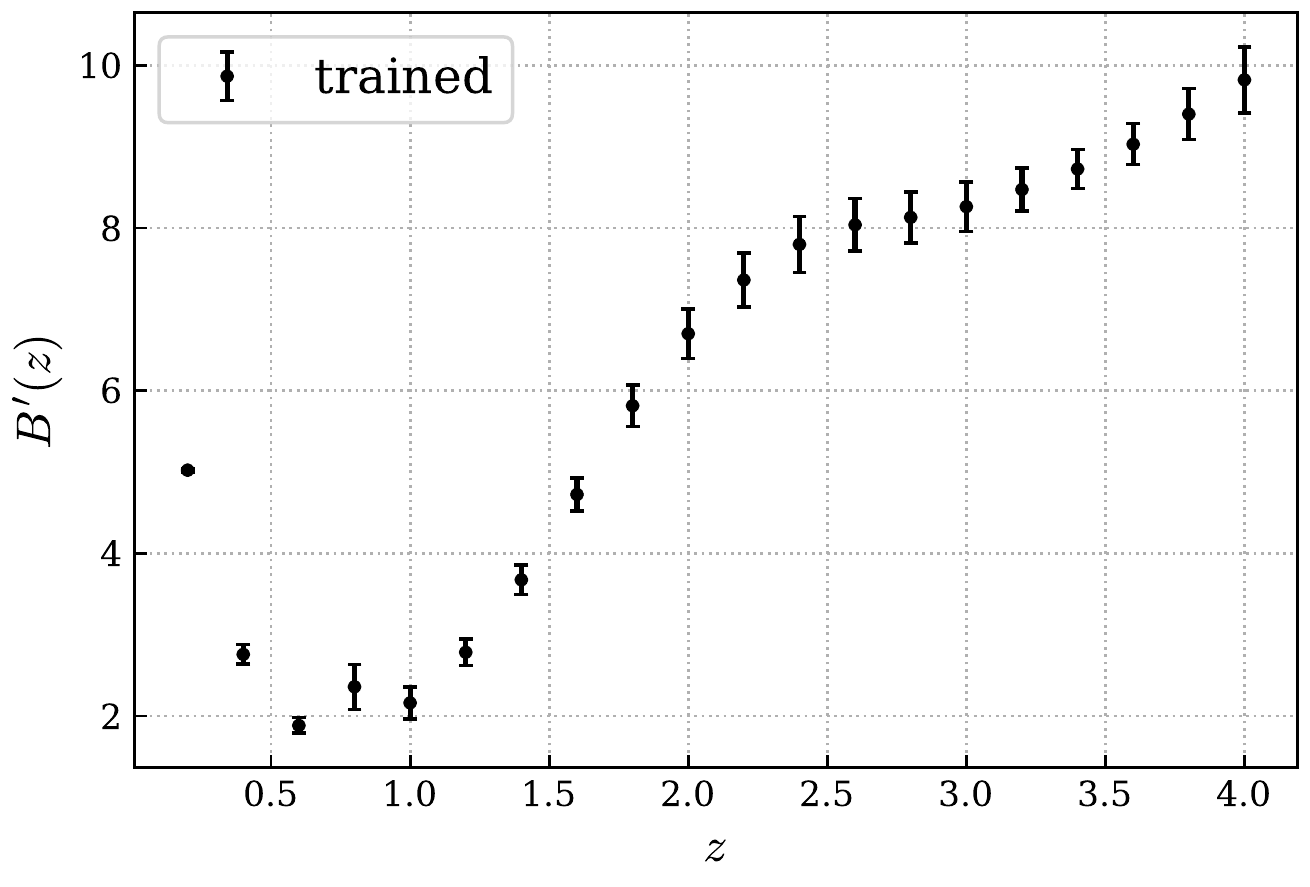}\label{fig:avmetrho}
			  }
			  \hspace{5mm}
			\subfigure[]
			 {\includegraphics[scale=0.6]{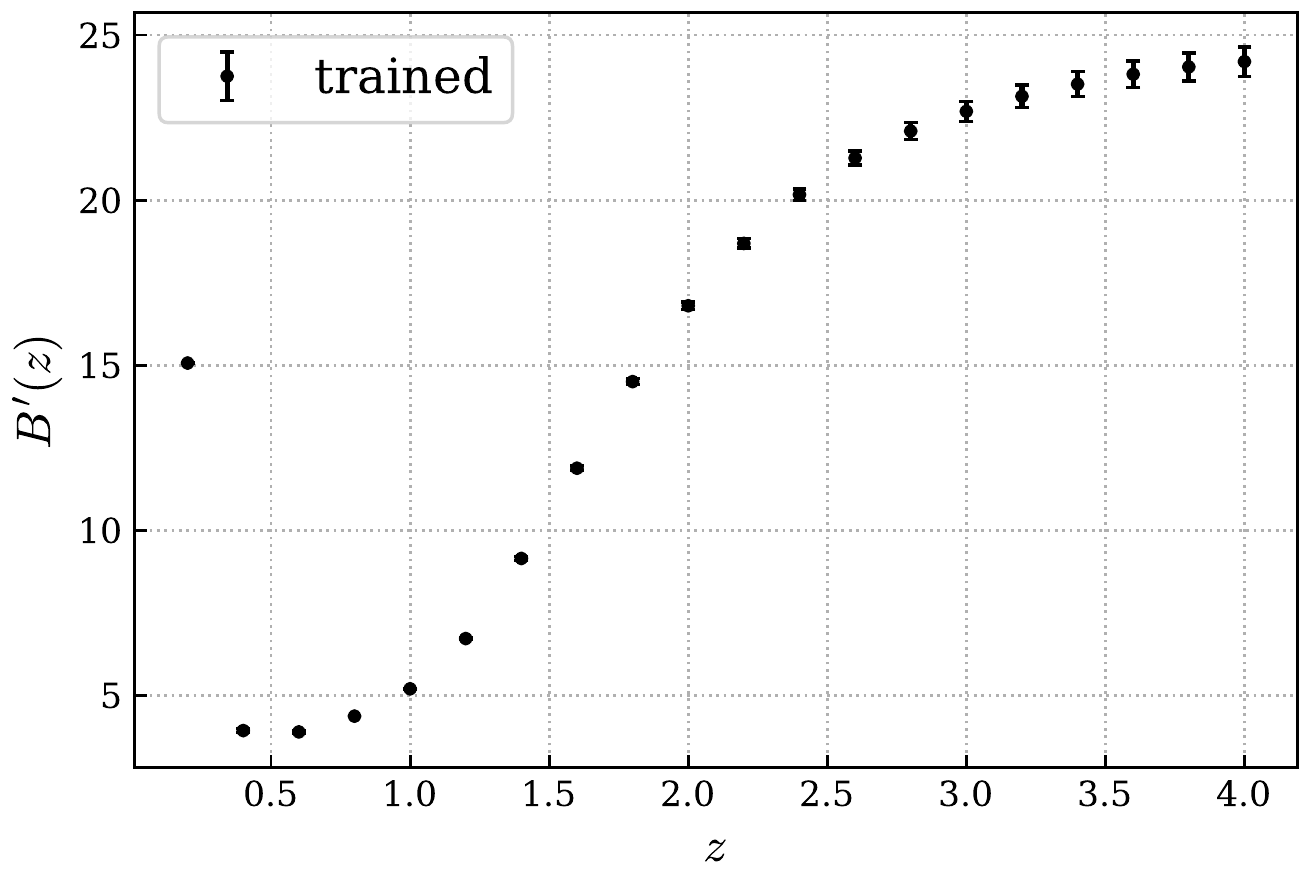}\label{fig:avmeta2}
			  }
			 \caption{
			The trained result of $B'(z)$, for (a) $\rho$ meson and for (b) 
			$a_2$ meson. The training is repeated twenty times and five times, respectively.  
			The average values over the iteration are plotted. The error bars represent $2\sigma$.
			 }
			 \label{fig:avmetrhoa2}
		\end{figure*}


\subsection{Model determined by experimental data}
\label{sec:exp}

Finally, we come to the actual aim of the deep learning method: the determination of the metric/dilaton functions
by the experimental data of the meson spectra.
We use the spectra of the $\rho$ meson ($S=1$) and the $a_2$ meson ($S=2$). 
Combining the results of these two, we can obtain $\Phi(z)$ and $A(z)$ separately, as shown in 
\eqref{eq:spin dependence of B}.

The experimental data of the tower of the $\rho$ meson spectrum \cite{Tanabashi:2018oca} is given as
$\rho(770)$, $\rho(1450)$, $\rho(1700)$, $\cdots$,  and we use only the first two levels, since it is expected by the analysis of 
the model of Ref.~\cite{Karch:2006pv} that the discretization errors are difficult to be handled for higher excitations. 
As for the $a_2$ meson, $a_2(1320)$, $a_2(1700)$ are reported as the established poles \cite{Tanabashi:2018oca}. 
We generate the input data with these values, and design our neural network. (See FIg.~\ref{fig:dataset_rho}
and Fig.~\ref{fig:dataset_a2} for the generated input data, and TABLE~\ref{tab:hyper parameter} for the hyperparameters 
of the architecture. For details of the hyperparameters, the initial weights, and  the dataset generation, see App.~\ref{app:hyperparameters}, App.~\ref{app:weight}, and App.~\ref{app:dataset}, respectively.)

	\begin{table}[h]
	\centering
	\begin{tabular}{|c||c|c|c|c|c|c|c|} \hline
        meson & $\Delta z$ & $N$ & $\epsilon$ & mass(GeV) &  num.~of input data \\ \hline \hline
        $\rho$ & 0.2 & 20 & 0.25 & 0.77, 1.45 & pos:2000/neg:2000 \\ \hline
				$a_2$ & 0.05 & 80 & 0.25 & 1.32, 1.70 & pos:3000/neg:3000 \\ \hline
	\end{tabular}
	\caption{Hyperparameters and dataset we use. The values of the mass spectra are taken from Ref.~\cite{Tanabashi:2018oca}.}
	\label{tab:hyper parameter}
	\end{table}

In the implementation of the dimension-ful quantities (such as the masses) into the numerical experiment,
we normalized everything appearing in \eqref{eq:eom of bulk field} in the unit of the AdS radius $L$.
It is actually the unique dimension-ful parameter in the simple AdS/QCD model \eqref{eq:effective action of vector sector}.
The input values in Fig.~\ref{fig:dataset_rho} and Fig.~\ref{fig:dataset_a2} are multiplied by $L$ when they are used in
the training.
The value of $L$ can be chosen arbitrarily, and here we choose it in such a way that 
the mass of the lowest $\rho$ meson (0.77GeV) is equal to the mass of the ground state ($m_n^2=4(n+1)$) 
of the model of Ref.~\cite{Karch:2006pv} in the unit of $L$, for simplicity.
This results in $L = \sqrt{4/0.77^2}  \sim 2.6$GeV${}^{-1}$ as the unit of length.

We repeat the training of the neural networks with those dataset,  twenty times for the $\rho$ meson and five times for the 
$a_2$ meson, separately. The network is optimized, and the emergent $B'(z)$s are obtained, which are shown in FIG.~\ref{fig:avmetrhoa2}.

\begin{figure*}[tb]
			  \centering
			\subfigure[]
			 {\includegraphics[scale=0.6]{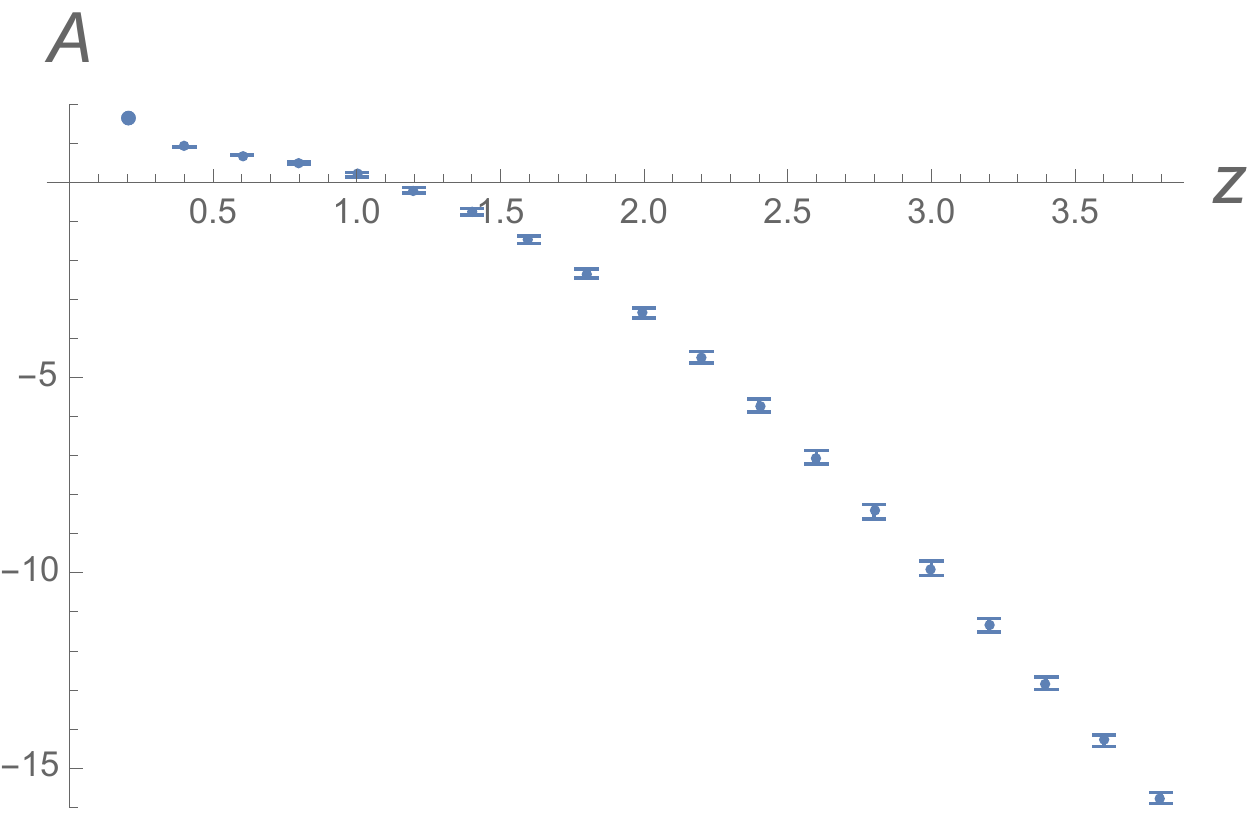}\label{fig:Aresult}
			  }
			  \hspace{4mm}
			\subfigure[]
			 {\includegraphics[scale=0.6]{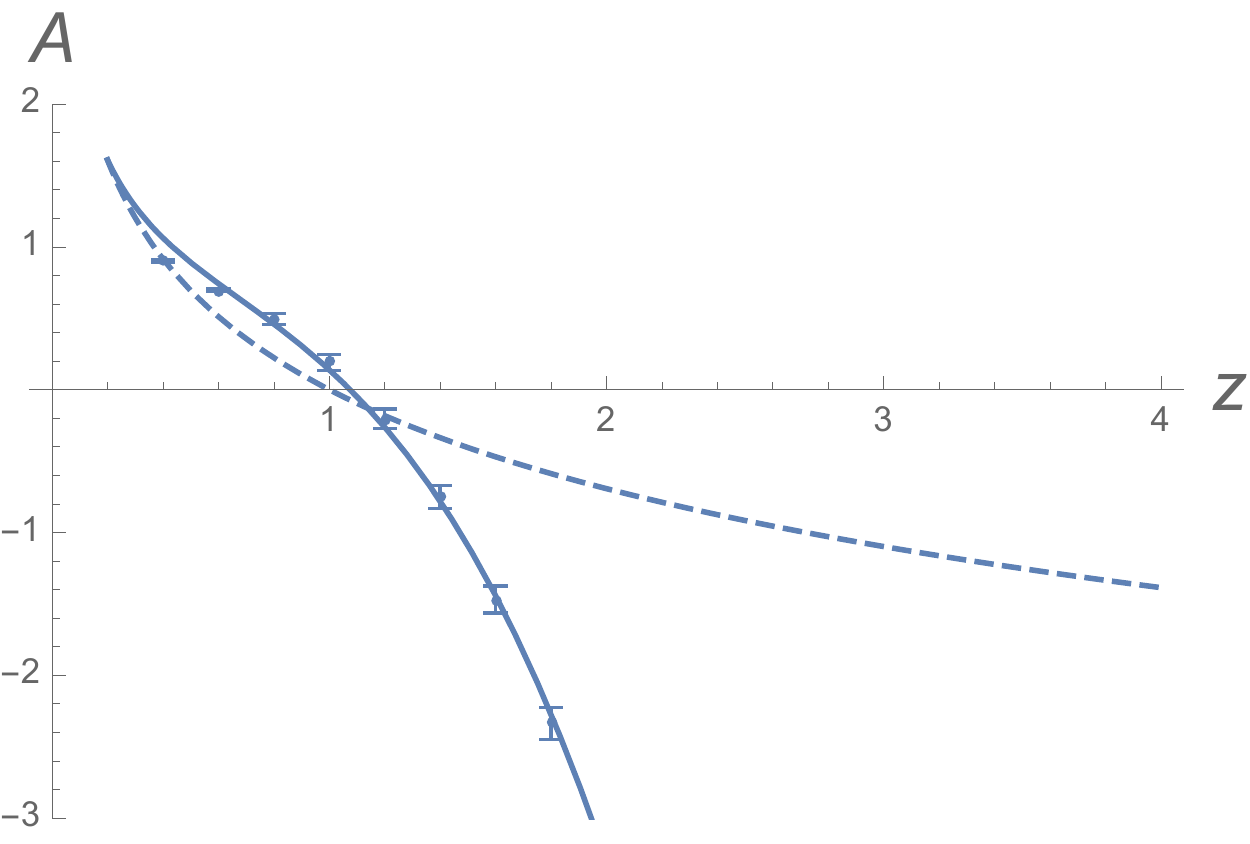}\label{fig:Acresult}
			  }
			 \caption{
			(a) Emergent metric function $A(z)$ with the statistical error bars. 
			(b) Enlarged figure near the boundary, $z\sim 0$. The solid line is the fitting function \eqref{Afit}. 
			The dashed line is $-\log(z)$ which is the AdS spacetime.
			 }
			 \label{fig:Ares}
		\end{figure*}

\begin{figure}[tb]
			  \centering
			 {\includegraphics[scale=0.6]{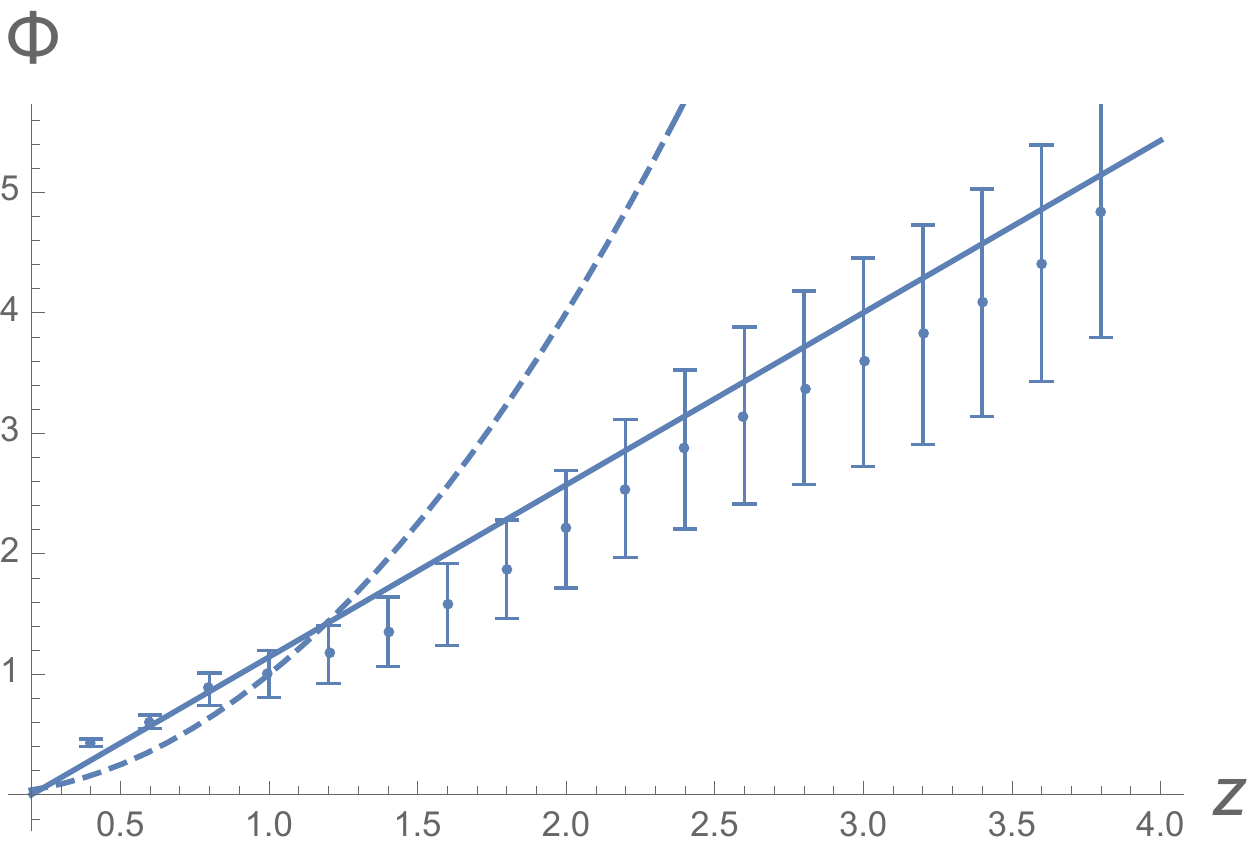}
			  }
			 \caption{
			Emergent dilaton profile. The solid line is a fit \eqref{Phifit}, and the dashed line is $\Phi=z^2$ of the model \cite{Karch:2006pv}.
			 }
			 \label{fig:Dres}
		\end{figure}

To obtain the metric and the dilaton, we calculate $B(z)$ by a discretized integration
		\begin{align}
		B(z=n\Dz)=\sum^n_{k=1}B'(k\Dz)\Dz+C \, . 
		\end{align}
We set the integration constant $C$ such that $B(\Dz)=\log \Dz$, 
because $B(z)$ should behave as an asymptotically AdS spacetime, $A(z) \simeq -\log z$ by the assumption.
From this integral we compute the metric profile function $A(z)$ and the dilaton profile function $\Phi(z)$. 
They are shown in FIG.~\ref{fig:Ares} and  FIG.~\ref{fig:Dres}.

The emergent metric $A(z)$ for the near-boundary region $0<z\leq 2.0$ 
can be consistently fit with a function
\begin{align}
A(z) = -\log z +a_1 (z\!-\!0.2) +a_2 (z\!-\!0.2)^2 + a_3(z\!-\!0.2)^3\, , 
\label{Afit}
\end{align}
with $a_1 = 0.849$, $a_2 = - 0.496$, and $a_3 = -0.433$. The first term $(-\log z)$ is for 
the AdS${}_5$ spacetime, and the corrections are obtained as above. 
The fitting Taylor series for the deviation from the AdS is in the power of $(z-0.2)$
where $z = \Dz = 0.2$ is the location of the initial layer.

The emergent dilaton profile 
is plotted in Fig.~\ref{fig:Dres}. The obtained points with the error bars
can be fit well with a linear function,
\begin{align}
\Phi(z) = \phi_1 (z-0.2)
\label{Phifit}
\end{align}
with a constant $\phi_1 = 1.43$. As one can see in Fig.~\ref{fig:Dres}, our deep learning have found
a linear dilaton profile, rather than the $z^2$ behavior which was originally anticipated in Ref.~\cite{Karch:2006pv}. 
The linear dilaton is a popular background in string theory as the world sheet theory is solvable, 
and it is interesting that it shows up in the machine learning.\footnote{In
the context of holographic QCD the exact linearity may cause a problem of continuous glueball spectra \cite{Gursoy:2007cb}.
Since we have not assumed a bulk gravity action which is necessary to argue the glueball spectra,
we leave this issue for our future problem.}

\subsection{Physical properties of the emergent spacetime}

With the optimized emergent metric and dilaton functions, we can even try to make a prediction. 
We calculate the mass of the next-higher level excitation of the $\rho$ meson and the $a_2$ meson.
Numerically calculating the eigenvalue of \eqref{eq:eom of bulk field} with our emergent $B(z)$, we summarize the result in TABLE~\ref{tab:model spectrum and exp spectrum}. 

		\begin{table}[bth]
			\centering
			\begin{tabular}{cccc}
				\hline \hline
				Meson & Model prediction  & Experiment \\
				& (GeV) & (GeV)  \\
				\hline
				$\rho$ & 1.52 & 1.70  \\
				$a_2$ & 2.44  & -  \\
				\hline \hline
			\end{tabular}
			\caption{Comparison of the experimentally measured spectra of the third-lowest 
			excitation of $\rho$ and $a_2$ \cite{Tanabashi:2018oca}, with our prediction 
			from the optimized AdS/QCD model on the emergent background.}
			\label{tab:model spectrum and exp spectrum}
		\end{table}

In the middle column of TABLE~\ref{tab:model spectrum and exp spectrum}, 
the calculated mass of the third-lowest excitation of each meson is shown. 
As for the $\rho$ meson, our optimized model predicts the mass which is compared with the experimental data
within 10 percent error. 
Our prediction of the $a_2$ meson mass will be confirmed in future experiment, 
as there is no established value in experiments
at present.

Our prediction accompanies a caution, as our architecture is based on various hyperparameters and discretization errors, 
as well as the fact that we only use the lowest and the second-lowest meson masses as the training data.
A change in widths introduced in
the positive/negative data in FIG.~\ref{fig:dataset} 
may cause unsuccessful trainings. One of the deficits of deep learning in general
is unknown relations between hyperparameters and generalization, and any prediction is with such a caution.
Nevertheless, we find that our prediction is still reasonable, which is encouraging.

\begin{figure}[tb]
			  \centering
			 {\includegraphics[scale=0.6]{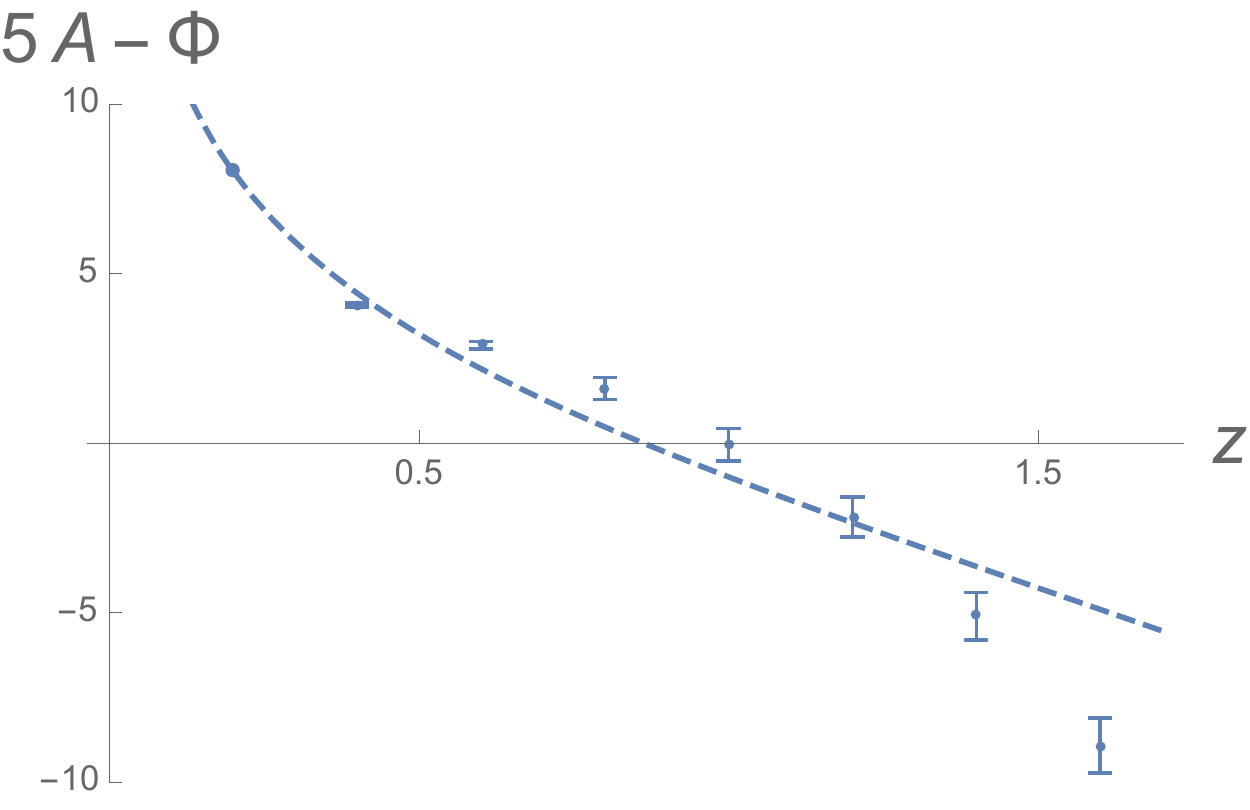}
			  }
			 \caption{
			The logarithm of the volume element with the dilaton, $\log[\sqrt{-g}e^{-\Phi}]=5A-\Phi$. The dashed line 
			is that of Ref.~\cite{Karch:2006pv}, $5A-\Phi=-5\log z - z^2$.
			 }
			 \label{fig:volume}
		\end{figure}

Finally, let us discuss the physical property of the emergent metric \eqref{Afit} and the dilaton \eqref{Phifit}.
Using these, we find an effective volume element $\sqrt{-g}e^{-\Phi}$ which is in the action 
\eqref{eq:effective action of vector sector}. Since the radial $z$-dependence of the volume element  reflects
physical properties of the geometry, we plot a logarithm of our optimized effective volume element $5A-\Phi$
in FIG.~\ref{fig:volume}. In the figure, for a comparison, we plot the dashed line which is $5A-\Phi$
of the model of Ref.~\cite{Karch:2006pv}.

We notice that when $z$ is increased and goes to the infrared, our volume element gradually 
deviates in the positive direction once, compared to the AdS spacetime of Ref.~\cite{Karch:2006pv}. 
This can be also seen from the fact that the 
first correction $a_1$ in \eqref{Afit} is positive, meaning that the AdS spacetime is deformed toward
a slower warping.
One possible interpretation is that this is a tendency toward a confining geometry,
which is consistent with that the non-supersymmetric
QCD is a confining theory at zero temperature.

Then, with a further increasing of $z$, our effective volume element goes below the AdS line.
The resultant bump of the volume element as a function of $z$ 
resembles the machine-learned geometry of the AdS/DL model for the lattice data of the 
QCD chiral condensate \cite{Hashimoto:2018bnb},
where at the finite temperature the emergent geometry consists of both the confining wall and the black hole horizon.
Our geometry is trained with experimental 
meson spectra, while the one in Ref.~\cite{Hashimoto:2018bnb} is with different QCD observables.
Finding a unique geometry consistent with more QCD observables is a challenging problem.

\section{Conclusion}
\label{sec:summary}

In this paper, we have proposed a deep learning architecture to discover an AdS/QCD model from a given experimental 
data of meson mass spectra. The neural network depth direction is identified with the emergent radial direction $z$
of the 5-dimensional model. The model is a soft-wall model based on Ref.~\cite{Karch:2006pv}, 
with the unknown metric function $A(z)$ of the curved geometry
and the unknown dilaton profile $\Phi(z)$. We have identified those profiles with weights of the deep neural network 
(FIG.~\ref{fig:illustration_DNN}).  By the supervised training with the lowest and the second-lowest $\rho$ and $a_2$ 
meson mass values as the training data,
our machine has found optimized profiles of the geometry $A(z)$ and the dilaton $\Phi(z)$ (FIGs.~\ref{fig:Ares} and \ref{fig:Dres}).
Therefore, the deep learning, based on the concept of AdS/DL \cite{Hashimoto:2018ftp,Hashimoto:2018bnb}, 
can derive an effective AdS/QCD model from QCD experimental data.

With the emergent geometry and the dilaton profile, we have calculated the excited meson masses. 
This prediction has turned out to be reasonable, although it should not be taken seriously, as 
the architecture has various regularization and discretization errors.
Nevertheless, the training to obtain the emergent geometry is worthwhile in a larger perspective.
This framework may open up a whole scheme of determining a better holographic model by a vast amount of data of QCD.
As we have emphasized in Sec.~\ref{sec:intro}, QFT has an infinite number of data, which are spectra and scattering amplitudes,
equivalent to $n$-point correlators of an infinite kinds of gauge invariant operators.
Finding better holographic models is equivalent to the feature extraction of QCD, which may help revealing the
hidden mechanism of how the AdS/CFT correspondence works.

Relatedly, the similarity between the holographic dictionary and the deep neural network architecture may have some 
physical origin, and such a standpoint may provide a new way to investigate two subjects which appear distantly related: 
quantum gravity and data science.
In the growing subject (see Ref.~\cite{Ruehle:2020jrk} for a recent summary of data science application to string theory),
the idea of equating a holographic spacetime with neural network 
\cite{Hashimoto:2018ftp,You:2017guh,Hashimoto:2018bnb,Vasseur:2018gfy,Hashimoto:2019bih,Hu:2019nea,Tan:2019czc}
may be intertwined with machine learning string landscapes initiated by 
Refs.~\cite{Krefl:2017yox,He:2017dia,Ruehle:2017mzq,Carifio:2017bov}.
Discovering a complete gravity dual of QCD is a challenging problem, and various data-scientific methods applied to
string theory may help for it.

\begin{acknowledgments}
We would like to thank Hong-Ye Hu and Yi-Zhuang You for valuable comments.
K.~H.~would also like to thank 
James Halverson, Andreas Karch, Elias Kiritsis, Sven Krippendorf, 
Francesco Nitti, Fabian Ruehle, Sotaro Sugishita, Akinori Tanaka, Akio Tomiya, 
and Ismail Zahed for valuable discussions.
The work of K.H.~is supported in part by JSPS KAKENHI Grant Number JP17H06462.

\end{acknowledgments}

\vspace*{15mm}


\appendix

\section{Supplement for coding our architecture}
\label{Sec:app}

In this appendix we provide details of our deep learning architecture.
Our Python code for the machine learning 
is written on the basis of the code used in Ref.~\cite{Hashimoto:2018bnb}.
The latter code is available at \cite{tanakaGithub} and 
we revise the forward function and the regularization terms of the code. 
In addition, the preparation of the input data shown in Fig.~\ref{fig:dataset} is necessary.
In the following, we provide details about the input layer, the hyperparameters, the loss function and the regularizations,
the initial weights, and the dataset generation.

\subsection{Input values at the initial layer}
\label{app:initiallayer}

At the end of Sec.~\ref{Sec:DNN}, 
we substitute the initial layer for units $v,~\pi$ with the asymptotic solution of a typical soft wall model as in 
\eqref{eq:BC on initial layer}, while the unit $m^2$ receives its value from the input data. We derive \eqref{eq:BC on initial layer}
in the following.

As mentioned, we assume that our model background is asymptotically AdS$_5$. 
This means that $A(z) \sim -\log z$ and $\Phi(z)$ vanishes at $z\sim0$. Then the function $B(z)$ defined in \eqref{eq:spin dependence of B} is approximated by $(2S-1)\log z$. Hence the bulk field equation \eqref{eq:eom of bulk field} reduces 
near the AdS boundary $z \sim 0$ to
\begin{align} \label{eq:eom of bulk field near the boundary}
	\partial_z \qty(\frac{1}{z^{2S-1}} \partial_z v)+ m^2 \frac{1}{z^{2S-1}} v = 0 \, .
\end{align}
Assuming a power-law configuration $v\sim z^\alpha$, then the equation above reduces to
\begin{align}
	&\alpha\qty(\alpha-2S)z^{\alpha-1-2S} + m^2 z^{\alpha+1-2S} = 0 \, , 
\end{align}
which forces us to choose $\alpha=0,2S$.
We can write a general solution as
\begin{align}
	v = a + bz^{2S} \, .
\end{align}
According to the AdS/CFT dictionary, the non-normalizable part $a$ corresponds to a source in the boundary theory 
and the normalizable part $b$ corresponds to the expectation value of an operator associated with the source. 
Since we are interested in hadron spectra without the source, we set $a=0$, and moreover we can choose $b=1$ 
due to the linearity of equation \eqref{eq:eom of bulk field near the boundary}. 
Therefore we estimate the behavior of $v$ near the boundary as
\begin{align} \label{eq:asymptotic behavior}
	v \sim z^{2S} \qq{}(z\sim0)
\end{align}
which also derives $\pi \sim 2S z^{2S-1}$, that is \eqref{eq:BC on initial layer}.


\subsection{Hyperparameters}
\label{app:hyperparameters}

As we described briefly in Sec.~\ref{Sec:hyp},
our hyperparameters $N,\Dz,\epsilon$ in our neural network are fixed by carefully looking at discretization errors.
Here we describe how to choose the value of the hyperparameters.
We choose $\Dz$ first, then determine the others. For the evaluation of the discretization errors, we need some analytic 
solutions for comparison, and we adopt the fluctuation solutions of Ref.~\cite{Karch:2006pv}.

\begin{figure}[tbh]
	\centering
	\includegraphics[width=0.9\linewidth]{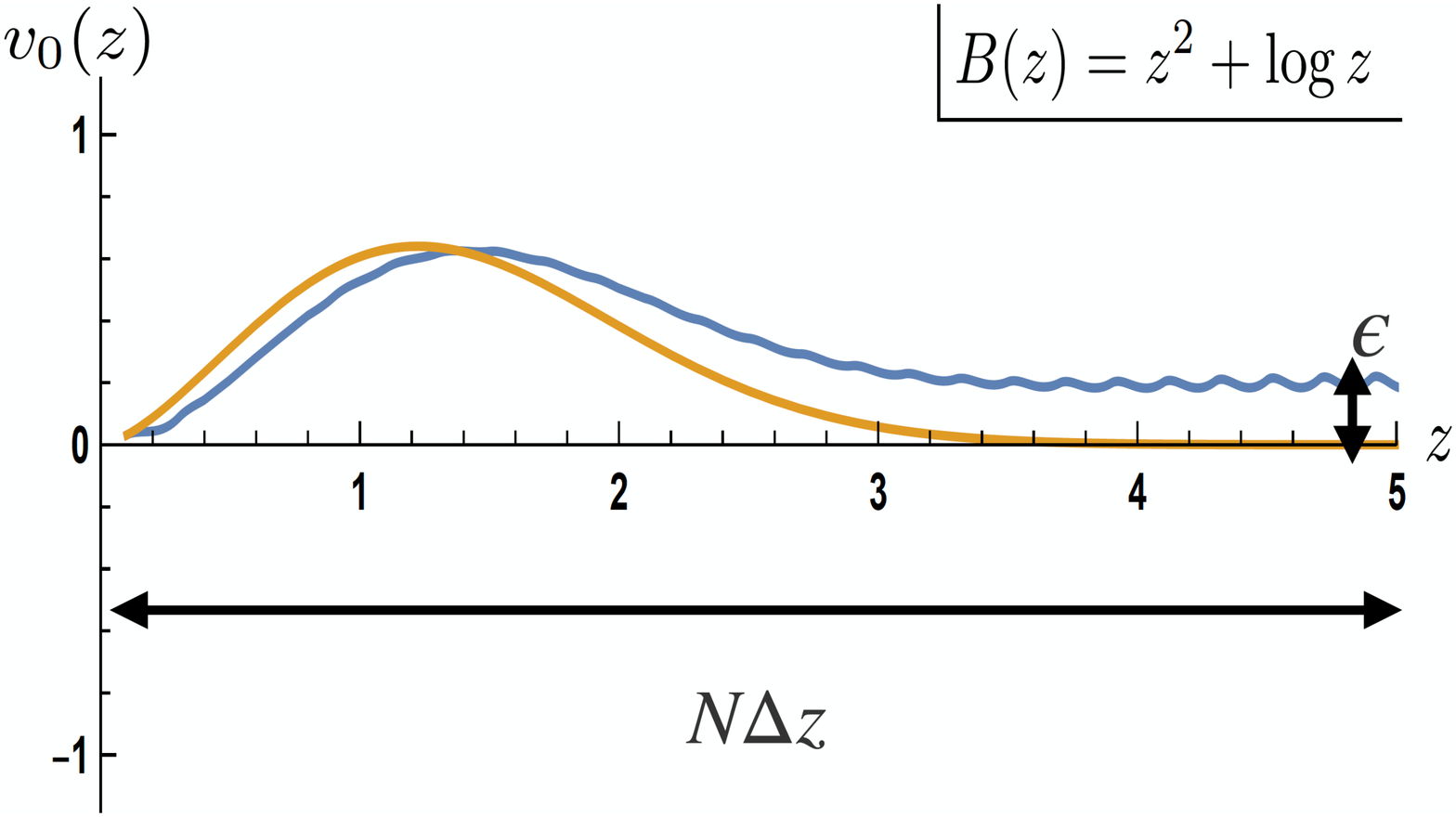}
	\caption{
		The ground sate solution of \eqref{eq:eom of bulk field} with the model background 
		given in Ref.~\cite{Karch:2006pv} ($B(z)=z^2+\log z$), solved analytically (yellow line) and numerically (blue line). 
		The blue line interpolates the plots of the numerical solution.
		}
	\label{fig:num. vs ana. in soft wall}
\end{figure}

FIg.~\ref{fig:num. vs ana. in soft wall} shows the lowest solution $v_0(z)$ 
of the equation of motion of bulk field, which is $v(z)$ with $m=m_{n=0}$
of the model \cite{Karch:2006pv}. Both the analytic solution and the numerical solution are plotted.
The numerical solution is obtained by Euler method with width $\Dz=0.2$. 
As mentioned in Sec.~\ref{Sec:hyp}, we have to look at the location where the solution converges, and also at 
the difference between the analytic solution and the numerical solution. 

First, the plot shows that the analytic $v_0$ almost touches the $z$ axis around $z=4$. 
This suggests to set $N=20$ in order to discriminate the normalizability of solution. Next, 
the numerical solution approaches 0.2, not 0, due to the discretization error. 
So we have to choose $\epsilon$ greater than 0.2. 
From the analysis above, we set these hyperparameters to the values 
shown in TABLE~\ref{tab:model spectrum and exp spectrum}.

Of course, we may choose a smaller $\Dz$ and a larger $N$. However, 
the discretization error does not change so much, while a larger $N$ makes the computation too heavy 
because the neural network becomes deeper. 

\subsection{Loss function and regularization}
\label{app:loss}
	
The total loss function we use is 
	\begin{align} \label{eq:loss}
	E' = E + \sum_{i} p_i \, , 
	\end{align}
where $E$ is an L1 loss function, which is commonly used in binary classification. 
Denoting $\vb*{y}'$ as the output of the neural network and  
$\vb*{y}$ as the output part of the training data, 
then the loss function is made of the L1 distance $\vb*{y}'-\vb*{y}$, 
	\begin{align}
		E\qty(\vb*{y}'-\vb*{y}) = \frac{1}{d} \sum_i^{d} \qty|y_i'-y_i| \, .
	\end{align}
Here $i$ labels the data and $d$ is the number of the data.
    
	The output of our neural network takes a binary value 0 or 1, which is transformed from the unit value of $v$ at the 
$N$th layer by its normalizability condition. 
This transformation is given by a smeared box (step) function which was also used in Ref.~\cite{Hashimoto:2018bnb}. 
For more details, see the function named ``t'' in the code of Ref.~\cite{Hashimoto:2018bnb}.

	The loss function \eqref{eq:loss} contains a sum of  $p_i$'s, which are the regularizations. 
	Explicitly, they have the following form:
	\begin{align} \label{eq:regularization terms}
		\begin{cases}
			& p_1 = \sum_{l=1}^N c_{1,l} \qty(\frac{2S-1}{l \Dz}-B'^{(l)})^2 \\
			& p_2 = \sum_{l=1}^N c_{2,l} \qty(B'^{(l)} - B'^{(l-1)})^2 \\
			& p_3 = \sum_{l=1}^N c_{3,l} \qty(B'^{(l)} - 2B'^{(l-1)} + B'^{(l-2)})^2  \\
			& p_4 = \sum_{l=1}^N c_{4,l} \, \mathrm{relu} \qty(B'^{(l-1)}-B'^{(l)} + \delta)
		\end{cases}
	\end{align}
Here $B'^{(l)}$ denotes the weight at $l$-th layer, or equivalently $B'(l\Dz)$. 
And the coefficient constant $c_{i,l}$ determines how their constraints are effective as the regularization. 
There are four kinds of regularizations, $p_i$ $(i=1,2,3,4)$. In the following we explain each in detail.

	$p_1$ is the regularization to require that the emergent background is asymptotically AdS. 
Note that $S$ is spin of meson whose mass spectrum we use as the input, and 
this factor comes from \eqref{eq:spin dependence of B}. 
Since this constraint is only for the boundary region of the 5-dimensional geometry,
we turn it on only at the first few layers. We set most of $c_{1,l}$ to 0. 
When we train the network with the $\rho$ meson spectrum,  we use the following coefficients:
	\begin{align}
		\{ c_{1,l} \} = \{ 0.001, 0.0001, 0.0001, 0, \cdots , 0 \} \, .
	\end{align}

	$p_2$ and $p_3$ are the regularizations to require that the emergent background is a smooth geometry. 
This constraint is necessary to be imposed over all the layers layer but the first few. 
So, in the $\rho$ meson case, we impose $p_2$ regularization on the 4th layer and deeper layers.
For $p_3$, after some tuning, we find that it is enough to impose it from the 7th layer. 
The coefficients used in the $\rho$ meson training is shown below.
	\begin{align}
		& \{ c_{2,l} \} = \{ 0, 0, 0, 0.00001, 0.00001, 0.00001, 0.0001, \cdots \} \, , \nonumber \\
		& \{ c_{3,l} \} = \{ 0, 0, 0, 0, 0, 0, 0.0001, \cdots \} \, .
	\end{align}	

	$p_4$ is the regularization to require that the weight $B'(z)$ is a monotonously increasing function. 
``relu'' is a built-in function of machine learning package, which is defined as
	\begin{align}
		\text{relu}(x) = 
		\begin{cases}
			& 0 \qq{for} x\le0 \\
			& x \qq{for} x>0
		\end{cases}
	\end{align}
Hence, if $B'$ decreases along two layers, this regularization provide a positive loss. 
A small constant $\delta$ plays a roll to require that $B'$ should increase at least 
by $\delta$. A nonzero small $\delta$ is necessary to train the $a_2$ meson data in our case.
The $p_4$ regularization is to require that 
$B(z)$ should provide the ``wall''-like behavior not to have a continuum spectra. For that,
we need this $p_4$ regularization at the large $z$ region, 
otherwise the model allows some unexpected continuous spectrum at a larger mass.
The coefficient used in the $\rho$ meson training is the same as that of $p_3$, 
	\begin{align}
		\{ c_{4,l} \} = \{ 0, 0, 0, 0, 0, 0, 0.01, \cdots \} \, .
	\end{align}

All the coefficients above is for the $\rho$ meson training. For the $a_2$ meson training, 
we choose the coefficients as follows:
	\begin{align}
		 & c_{1,1}  = 0.01 \, , 
		 \nonumber \\
		 &c_{1,i}  = 0.001 \,(i=2,\cdots,8)\, , 
\nonumber \\
		 &c_{3,i}  = 0.01\,(i=9,\cdots,12) \, , 
\nonumber \\
&c_{3,j}  = 0.005\,(j=13,\cdots,21) \, , 
\nonumber \\
&		 c_{3,k}  = 0.001\,(k=22,\cdots,80) \, , 
\nonumber \\
&		 c_{4,i}  = 10\, (i=61,\cdots,80) \, ,
	\end{align}
and the other components are put to zero.


\subsection{Initial weight}
\label{app:weight}

	For a successful training, we need to control the initial values of the weights to some extent. 
In the $\rho$ meson training, we initially sample the values of the weights at the $l$-th layer from the normal distribution 
whose mean is $1+0.5 l$ with the standard deviation $0.3l$ when we initialize the neural network.
In the $a_2$ meson training, we initially sample them from the normal distribution whose mean is $3/(0.5l+0.05)+0.05l-7$ with 
the standard deviation $10$.


\subsection{Training dataset}
\label{app:dataset}

	Here we present our Python code to generate the dataset given in Fig.~\ref{fig:dataset_rho} and Fig.~\ref{fig:dataset_a2}.
Our training trials observe that the success of the training depends on the range of the positive data values
and also the local density of data points, which is presented explicitly 
in the following codes for the $\rho$ meson and for the $a_2$ meson. 

\vspace{3mm}



\begin{widetext}

\noindent
1) Dataset for $\rho$ meson (Fig.~\ref{fig:dataset_rho})
	\begin{lstlisting}
		data_size = 1000
		wd = 0.1

		o_pos_data = []
		F_pos_data = []

		o_neg_data = []
		F_neg_data = []

		omega=[0.77,1.45] # mass[GeV]

		while True:
			o = np.random.uniform(0, 1.65)# sampling from uniform distribution.
			list_delta_o = []
			for i in range(len(omega)):
				list_delta_o.append(abs(o - omega[i]))
			delta_o = min(list_delta_o)
			if delta_o < wd and len(F_pos_data) < data_size:
				o_pos_data.append([o])
				F_pos_data.append([0])
			elif delta_o >= wd and len(F_neg_data)<data_size:
				o_neg_data.append([o])
				F_neg_data.append([1])
			if len(F_pos_data)==data_size:
				break

		while True: # increase neg_data between m_0 and m_1, and greater than m_1
			o = np.random.uniform(1., 1.2)# sampling from uniform distribution.
			o_ = np.random.uniform(1.6,1.65)
			o_neg_data.append([o])
			F_neg_data.append([1])
			o_neg_data.append([o_])
			F_neg_data.append([1])
			if len(F_neg_data)==data_size*4:
				break

		while True: # increase pos_data around m_0
			o = np.random.uniform(omega[0]-wd,omega[0]+wd)# sampling from uniform distribution.
			o_pos_data.append([o])
			F_pos_data.append([0])
			if len(F_pos_data)==data_size*2.5:
				break

		while True: # increase pos_data around m_1
			o = np.random.uniform(omega[1]-wd,omega[1]+wd)# sampling from uniform distribution.
			o_pos_data.append([o])
			F_pos_data.append([0])
			if len(F_pos_data)==data_size*4:
        break
	\end{lstlisting}

\vspace{3mm}

\noindent
2) Dataset for $a_2$ meson (Fig.~\ref{fig:dataset_a2})
	\begin{lstlisting}
		data_size = 500
		wd = 0.05

		o_pos_data = []
		F_pos_data = []

		o_neg_data = []
		F_neg_data = []

		omega=[1.32,1.70] # mass[GeV]

		while True:
			o = np.random.uniform(0, 2.0)# sampling from uniform distribution.
			list_delta_o = []
			for i in range(len(omega)):
				list_delta_o.append(abs(o - omega[i]))
			delta_o = min(list_delta_o)
			if delta_o < wd and len(F_pos_data) < data_size:
				o_pos_data.append([o])
				F_pos_data.append([0])
			elif delta_o >= wd and len(F_neg_data)<data_size:
				o_neg_data.append([o])
				F_neg_data.append([1])
			if len(F_pos_data)==data_size:
				break

		while True:
			o = np.random.uniform(1.8, 2.0)# increase neg_data greater than m_1
			lo_neg_data.append([o])
		        F_neg_data.append([1])
			if len(F_neg_data)==data_size*2.5:
				break


		while True:
			o = np.random.uniform(1.4, 1.6)# increase neg_data between m_0 and m_1
			o_neg_data.append([o])
			F_neg_data.append([1])
			if len(F_neg_data)==data_size*6:
				break

		while True:
			o = np.random.uniform(1.30, 1.34)# increase pos_data around m_0
			o_pos_data.append([o])
			F_pos_data.append([0])
			if len(F_pos_data)==data_size*3.5:
				break
				
		while True:
			o = np.random.uniform(1.68, 1.72)# increase pos_data around m_1
			o_pos_data.append([o])
			F_pos_data.append([0])
			if len(F_pos_data)==data_size*6:
				break				
				
	\end{lstlisting}

\end{widetext}



\end{document}